\documentstyle[aps,epsfig]{revtex}

\def\Journal#1#2#3#4{{#1} {\bf #2}, #3 (#4)}

\def\NPB{{\em Nucl. Phys.} B}
\def\NPA{{\em Nucl. Phys.} A}
\def\PLB{{\em Phys. Lett.} B}
\def\PRL{\em Phys. Rev. Lett.}
\def\PRD{{\em Phys. Rev.} D}
\def\PRC{{\em Phys. Rev.} C}

\begin{document}

\preprint{\vbox {\hspace*{\fill} DOE/ER/40762-156\\
          \hspace*{\fill} UMD PP\#99-011}} 
\vspace{.5in}

\title {Low Energy Theorems For Nucleon-Nucleon Scattering}

\author{Thomas D. Cohen}

\address{Department of 
Physics, University of~Maryland, College~Park, MD~20742-4111}

\author{James M. Hansen}

\address{Montgomery Blair  High School,  Silver Spring, MD 20901}

\maketitle

\vspace{.25in}

\begin{abstract}

Low energy theorems are derived for the coefficients of the effective
range expansion in s-wave nucleon-nucleon scattering valid to leading
order in an expansion in which both $m_\pi$ and $1/a$ (where $a$ is
the scattering length) are treated as small mass scales.  Comparisons
with phase shift data, however, reveal a pattern of gross violations
of the theorems for all coefficients in both the ${}^1S_0$ and
${}^3S_1$ channels.  Analogous theorems are developed for the energy
dependence $\epsilon$ parameter which describes ${}^3S_1$--${}^3D_1$
mixing.  These theorems are also violated.  These failures strongly
suggest that the physical value of $m_\pi$ is too large for the
chiral expansion to be valid in this context.  Comparisons of $m_\pi$
with phenomenological scales known to arise in
the two-nucleon problem support this conjecture. \vspace{.5in}
\end{abstract}

\section{Introduction \label{intro}}

Chiral perturbation theory ($\chi$PT) has  proven to be an exceptionally 
powerful tool
for analyzing interactions involving pions in hadronic physics. As noted by
Weinberg \cite{Weinberg1}, however, the implementation of conventional 
$\chi$PT in  low energy nuclear physics is problematic since the
s-wave scattering lengths are  important distance scales which are much
larger than $m_\pi$.
This ``unnatural 
scattering length'' problem has received considerable attention
during the past several years
\cite{Weinberg1,KoMany,Parka,KSWa,CoKoM,DBK,cohena,Scal,Fria,Sa96,LMa,GPLa,Adhik,RBMa,Bvk,aleph,Parkb,Gegelia1,Gegelia2,steelea,KSW,KSW2,CGSS,cohenb,Sav,CH,steeleb}.
 The goal of this work   is to implement the ideas of
effective field theory (EFT)\cite{EFT} to a problem in which both $m_\pi$ and
$1/a$ are light scales.  Much of the  power of EFTs stems from the fact that
the calculation of observables is formulated in terms of an expansion parameter
so that one has an {\it a priori} estimate of the accuracy of the calculation.
However, many of the  attempts to deal with the ``unnatural
scattering length'' problem, beginning with Weinberg's\cite{Weinberg1},
formulate the expansion at the level of a potential (or two-particle 
irreducible kernel)  rather than the physical
observables.  In such a formulation, one has no immediate
estimate of the accuracy of particular observables.
 Recently a scheme was introduced in which observables can be
expressed in terms of a consistent power counting scheme\cite{KSW,CH}. 
In principle, such an approach realizes the full power of the EFT idea.  

The  essence of this scheme is power counting in a single scale which we
denote as  $Q$:
\begin{equation}
m_\pi \sim   Q  \; \; \; 
1/a \sim  Q \; \; \; 
k \sim  Q
\label{pc}
\end{equation}
where $a$ is an s-wave scattering length, and $k$ an external momenta.
All other scales in the problem are assumed to be heavy and will be denoted 
$\Lambda$.
The power counting implies  that the external 
momenta must be small in the sense that $k/ \Lambda \ll 1$; however,
 one must include
all orders in $k a$ and $k/m_\pi$.  The inclusion of $k a$ and $k/m_\pi$
 to all orders introduces 
important nonperturbative effects.  This power counting scheme has been
implemented using dimensional regularization with the PDS scheme \cite{KSW}.
Alternatively, one may implement it directly in configuration space using a
cutoff \cite{CH}. In the case of s-wave scattering,
it was shown that the
two implementations are completely equivalent at next-to-leading order\cite{CH}
and it is plausible that this equivalence holds generally. We  refer to any scheme
which systematically implements eq.~(\ref{pc}) in the calculation of
observables as the $Q$ counting scheme.
A number of observables have been calculated using $Q$ counting at low order: 
N-N scattering\cite{KSW}, deuteron electro-magnetic form factors \cite{KSW2}
and deuteron polarizabilities \cite{CGSS}.  All of these 
calculations appear to  
describe the experimental data reasonably well.

In this paper we use $Q$ counting   to derive low energy theorems for
coefficients of the effective range expansion (ERE).  The ERE is a useful way 
to parameterize s-wave scattering when the scattering length is unnaturally
long.  It is given by
\begin{equation} 
k \cot (\delta) \, = \, -\frac{1}{a} \, + \, \frac{1}{2} r_e \,k^2 \, + \, v_2 \,  k^4 +
\, v_3 \, k^6 + \,v_4 \, k^8 + \ldots \label{effrangeexp}
\end{equation}
As we will show below, the $v_i$ coefficients ( $i \ge 2$ ) can be calculated in a
next-to-leading order calculation of $k \cot (\delta)$.  One obtains
predictions which are valid up to corrections of relative order $Q/\Lambda$,
where $\Lambda$ is the scale characterizing the short-ranged physics.
Moreover, these are direct predictions with no free parameters; in this
sense these are low energy theorems. However, when we compare these 
predicted $v_i$ coefficients with ones extracted from a fit to the scattering
data, we find that they fail quite badly; they predict $v_i$'s several times
larger than those extracted from the data.    We also derive theorems, valid to corrections of
relative order $Q/\Lambda$, for
coefficients characterizing the momentum dependence of ${}^3S_0$--${}^3D_0$ 
mixing at low momentum.  These theorems also do not work well; they
predict coefficients which differ from the those extracted from data by more than
100\%.

This poses  an interesting question.  Why are these theorems failing
so badly when
other predictions of $Q$ counting seem to work rather well?  One likely reason
is that the successful predictions of the $Q$ counting depend on 
different physics than what is being tested by these low energy  theorems.
The present predictions are extremely sensitive to the role played by the 
pion exchange in the two nucleon s-wave channel; most of the high quality predictions are not.
It is useful to note that the $Q$ counting EFT program of refs. \cite{KSW,CH}, 
has  two main components---i) 
dealing with  the ``unnaturally long scattering length'' problem 
($Q \sim 1/a$),
and ii) ``including pions'' ($Q \sim m_\pi$)\cite{Sav}. 
It is important to test both.   
 In $Q$ counting $1/a$ and $m_\pi$ are both formally treated as being of order
$Q$.  In reality, however,  $m_\pi \gg 1/a$ in both the singlet and triplet 
channels.  Accordingly it is  possible that most of the success seen to date
is coming about because
the expansion in $1/a$ is working even if the expansion in $m_\pi$ is failing.

 The importance of testing the ``including pions''
part of the program should be very clear.
Much of the original motivation for developing the effective
field theory approach for nuclear scattering  was to exploit chiral symmetry 
and develop a description
of nuclear phenomenon in terms of a controlled chiral
expansion.  The advantage of   including pions as explicit light degrees of 
freedom,
rather than integrating them out as heavy, should be obvious.  If the chiral
expansion is under control, the inclusion of explicit pionic degrees of freedom
 will significantly
improve the predictions at low $k$ and will significantly increase the maximum
value of $k$ for which the effective treatment is useful. This suggests a
simple way to test the ``including pions'' part of the $Q$ counting program.
One should calculate qualities in two EFTs---one including explicit pions and
implementing the $Q$ counting in eq.~(\ref{pc}), and the other in an EFT with
pions integrated out and using $Q \sim k, 1/a$ as the basis for the power
counting. The differences between these two calculations (done at the same
order in $Q$) is a measure of the effect of ``including pions''.

For many observables, the pion integrated out theory may give  extremely 
accurate predictions.  Both  the quartet
nucleon-deuteron scattering length \cite{Bvk} and the deuteron charge radius
\cite{cohenb} were predicted using systematic EFT approaches with the pion 
integrated out.  
Clearly the two preceding successes reflect the usefulness of the $Q \sim k,
1/a$ counting scheme (which is essentially the
$\aleph$ counting introduced by van Kolck\cite{aleph}),  
and have nothing to do with chiral counting. 
It is possible that most of the successes of the $Q$ counting approach
similarly do not test the pionic aspects. 
Accordingly before concluding that chiral perturbation theory aspect
of $Q$ counting in nuclear
physics is under control, it is critical to find observables which  test the
chiral physics. 

The $v_i$ coefficients in the  effective range expansion are an ideal way to 
test the pion physics.  The central reason for  using the effective range expansion
is to isolate the effects of an unnaturally small scattering length from the 
remaining terms in the  expansion.  Consider first a theory with pions
integrated out.  In such a theory, the $v_i$ coefficients must be
${\cal{O}}(Q^0)$ since they are insensitive to $a$ and there are no other light
scales in the problem.  In contrast, as we will explicitly demonstrate below,
these coefficients are non-analytic in $Q$ with $v_j \sim Q^{- 2 j + 2}$. 
Thus, they are large in a theory with explicit pions and order unity in a theory
with pions integrated out.  Since there is a large difference between the two
cases the predictions for the $v_i$'s sensitively test the pion physics.
We will also show below that when pions are explicitly included,
 the calculated $v_i$ are not merely non-analytic in $Q$. 
 For all of the $v_i$, each  of the leading order
terms in $Q$ are  non-analytic in $m_\pi$; diverging as $m_\pi \rightarrow 0$.
Since these coefficients diverge as $m_\pi \rightarrow 0$
it is clear that they are dominated by pionic effects.

This paper is organized as follows:  In the following section we derive low
energy theorems for the  coefficients in the effective range expansion; 
analogous theorems for ${}^3S_1$--${}^3D_1$ mixing  are derived next.  
After that we compare the predictions with coefficients
extracted from the Nijmegen partial wave analysis \cite{Nij} of the scattering  data.
  Finally  we discuss 
why these failures might have been anticipated in light of the known scales 
in the nucleon-nucleon problem, and  how the failures
of the low energy theorems  can be reconciled with the apparently
successful   predictions for the phase shifts reported in ref. \cite{KSW}.

\section{Low Energy Theorems for the Effective Range Expansion}

In developing physical intuition for our low energy theorems, we find it 
somewhat more useful to use the cutoff formulation discussed in ref. \cite{CH}.
The essential physical idea in this approach is to implement the separation 
of long distance physics from short distance physics directly in configuration 
space.  A radius, $R$, is introduced as a matching point between long and short distance
effects; renormalization group invariance requires that physical quantities 
must  be independent of $R$.  It is important, however, that $R$ be chosen
large enough so that essentially all of the effects of the short distance
physics is contained within $R$. The potential is divided into the sum of two pieces,
a short distance part which vanishes for $r>R$ and long distance part which vanishes for 
$r<R$. 
At $R$, the information about short 
distance effects is entirely contained in the energy dependence of the 
logarithmic derivative (with respect to position) of the wave function at $R$.
Thus, provided we can  parameterize this information systematically, we can formulate
the problem in a way which is insensitive to the details of the short distance
part of the potential.
For $r>R$ the Schr\"odinger equation is solved subject to the boundary 
conditions at $R$. 
For s wave scattering, the wave function at $R$ may be parameterize as 
$A \sin (k r + \delta_0)$; the energy dependence of the logarithmic derivative
is independent of $A$ and can be expressed in terms of an expansion similar to 
an effective range expansion:
\begin{equation} 
k \cot (\delta_0) \, = \, -1/a_{\rm short} \, + \, 1/2 \,r_e^0\,k^2 \, + \, v_2^0 \,  k^4 +
\, v_3^0 \, k^6 + \,v_4^0 \, k^8 + \ldots \label{effrange0}
\end{equation}
Power counting in $Q$ for s wave scattering can be implemented straightforwardly.
All of the coefficients in the preceeding expansion are assumed to be order $Q^0$
except the first term ($-1/a_{\rm short}$ ) which will be taken to be order $Q^1$
to reflect the unnaturally large scale  of the scattering length.  Power counting
for the long range part of the potential simply follows Wienberg's analysis
\cite{Weinberg1}, with the previso that the potentials are only used for $r >R$.
At order $Q^2$ in $k \cot (\delta)$, only the simple one pion exchange contribution
to the $V_{\rm long}$ contributes.  The power counting also justifies an iterative
solution of the Schr\"dinger equation for $r>R$ along the lines of a conventional 
Born series.  It differs from the usual Born series in that the boundary conditions
at $R$ are implemented.  Finally, $Q$ counting is used in expanding out the final
expression for $k \cot (\delta)$.
  
Carrying out this program gives the following expression for $k \cot (\delta )$ 
at order $Q^2$ for the ${}^1S_0$ channel 
\begin{eqnarray}
k \cot (\delta) & = &-\frac{1}{a_0} \, + \, m_\pi^2 \, \left[d + \,
 \frac{g_A^2  M}{16 \pi f_\pi^2} \, \left (\gamma + \ln (m_\pi R) \right) \right ]
 \, + \, \frac{1}{2} \, r_e^0 \,k^2 
\, - \,  \frac{1}{a_0^2} \, \frac{g_A^2  M}{64 \pi f_\pi^2} \,
\left( \frac{m_\pi^2}{k^2} \right )
\ln \left
(1 + \frac{4 k^2}{m_\pi^2} \right ) \nonumber \\ \nonumber \\
 & + &  \, \frac{m_\pi}{a_0} \, \frac{g_A^2  M}{16 \pi f_\pi^2} \, 
 \left( \frac{m_\pi}{k} \right )\,
 \tan^{-1} \left ( \frac{2 k}{m_\pi} \right ) \,
 + \, m_\pi^2 \, \frac{g_A^2  M}{64 \pi f_\pi^2} \, 
 \ln \left (1 + \frac{4 k^2}{m_\pi^2} \right ) 
 \label{kcotd1}\end{eqnarray}
We use the convention in which $f_\pi$ = 93 MeV.  
Apart from well-known parameters from pionic physics, there are three
parameters---$a_0$, $d$ and $r_e^0$.  Where   
and $1/a_{\rm short}$ from eq.~(\ref{effrange0}) is rewritten as 
$1/a_0 + d m_{\pi}^2$.
 with $1/a_0 \sim Q$ and $d m_\pi^2 \sim Q^2$.
 These parameters fix the energy dependence of the
wave function at the matching scale $R$; renormalization group invariance
 requires $d$ to depend on R logarithmically.
The parameter $a_0$ corresponds to the scattering length at lowest order in
the $Q$ expansion.  It 
 is related to the observed scattering length
by 
\begin{equation}
 - \frac{1}{a} \, = \, - \frac{1}{a_0} + \, m_\pi^2 \, \left[ d \, + \, 
 \frac{g_A^2  M}{16 \pi f_\pi^2} \, \left ( \gamma + \ln (m_\pi R) \right) \right ]
 \, + \, \frac{g_A^2  M}{16 \pi f_\pi^2} \, \left( 
 \frac{2 m_\pi}{a_0} \, - \, \frac{1}{a_0^2} \right )  \, =  \,  
 - \,\frac{1}{a_0} + {\cal O}(Q^2/\Lambda)
 \label{scat}
 \end{equation}

The PDS scheme is superficially quite different in mathematical detail but 
ultimately describes the same physics based on $Q$ counting.
In terms of the parameters in the PDS formulation, one obtains the same
expression\cite{CH} provided one relates the $C_0$, $D_2$ and $C_2$ coefficients
in terms of $a_0$, $d$ and $r_e^0$ according to 
$$\frac{4 \pi}{M} \, \frac{1}{-\mu + 1/a_0} \, = \, C_0 $$

$$\frac{1}{2} \, r_e^0 \,  =  \, \frac{C_2 M}{4 \pi } \, \left ( \mu^2 \, - \, 
 \frac{2 \mu}{a_0} \, + \, \frac{1}{a_0^2} \right ) $$
 
\begin{eqnarray}
m_\pi^2 \, \left[d \, + \, 
 \frac{g_A^2  M}{16 \pi f_\pi^2} \, \left (\gamma + \ln (m_\pi R) \right) \right] \, & = &
 \,
  \frac{g_A^2  M}{16 \pi f_\pi^2} \, \left ( m_\pi^2 \,   \ln
\left (\frac{m_\pi}{\mu} \right ) \, - \,
 m_\pi^2 \, + \, \frac{1}{a_0^2} \, - \, 2 \frac{\mu}{a_0}  \, + \mu^2 \right )
\nonumber \\ \nonumber\\
&  +  &\, \frac{D_2 M}{4 \pi} \left ( m_\pi^2 \mu^2 \, -  \, 
\frac{2 m_\pi^2 \mu}{a_0} \, + \, \frac{m_\pi^2}{a_0^2} \right)  
\label{equiv}\end{eqnarray}
There is a subtle issue associated with the equivalence given above concerning
the  behavior 
as the chiral limit is approached.  This issue is discussed in the appendix.

The parameter $a_0$ plays a critical role in eq.~(\ref{kcotd1}).  It not only
plays a dominant role in fixing the value of $k \cot (\delta)$ as $k \rightarrow
0$, it also appears in the $\tan^{-1}$ term and one of the $\ln$ terms which
have nontrivial dependence on $k/m_\pi$.  Unfortunately, we do not have
a direct experimental way to fix $a_0$.  On the other hand, from
eq.~(\ref{scat}) we see that $1/a_0 \,  = 1/a (1 + {\cal O}(Q/\Lambda))$.
 Accordingly, if we replace $1/a_0$ with $1/a$ in the terms of order $Q^2$
  in
 eq.~(\ref{kcotd1}),  any error made is order $Q^3$---which is beyond the order
 to which we work.  Thus, to order $Q^2$,   
 \begin{eqnarray}
k \cot (\delta) & = &-\frac{1}{a_0} \, + \, m_\pi^2 \, 
 \left [ d \, + \, \frac{g_A^2  M}{16 \pi f_\pi^2} \, \left ( \gamma + \ln (m_\pi R) \right) \right ]
 \, + \, \frac{1}{2} \, r_e^0 \,k^2 
\, - \,  \frac{1}{a^2} \, \frac{g_A^2  M}{64 \pi f_\pi^2} \,
\left( \frac{m_\pi^2}{k^2} \right )
\ln \left
(1 + \frac{4 k^2}{m_\pi^2} \right ) \nonumber \\ \nonumber \\
 & + &  \, \frac{m_\pi}{a} \, \frac{g_A^2  M}{16 \pi f_\pi^2} \, 
 \left( \frac{m_\pi}{k} \right )\,
 \tan^{-1} \left ( \frac{2 k}{m_\pi} \right ) \,
 + \, m_\pi^2 \, \frac{g_A^2  M}{64 \pi f_\pi^2} \, 
 \ln \left (1 + \frac{4 k^2}{m_\pi^2} \right ) 
 \label{kcotd2}\end{eqnarray}

The form of eq.~(\ref{kcotd2})
 also holds for the triplet channel---except for
changes in the value of the parameters.   
The distinction between the singlet and triplet
channel is only due to the tensor force.  The effect of the tensor force on the
s-wave requires at least two iterations of the pion exchange  and contributes to $k \cot
(\delta)$ only at order $Q^3$ and beyond\cite{KSW}. 

The predicted coefficients  in the effective range expansion are easily obtained.  One
begins with the expression for $k \cot (\delta)$ in eq.~(\ref{kcotd2}) and
simply makes a Taylor expansion with respect to $k^2$.  From the definition of
the effective range expansion in eq.~(\ref{effrangeexp}) the $v_i$ coefficient
is simply the coefficient in the Taylor expansion multiplying $k^{2i}$ (for $i
\ge 2$)  We  find:
\begin{eqnarray}
v_2 \, &  = & \, \frac{g_A^2 M}{16 \pi f_\pi^2} \, \left ( \, -\frac{16}{3 a^2
\,
m_\pi^4}\,  + \, \frac{32}{5 a \,m_\pi^3} \, - \,\frac{2}{m_\pi^2} \right
)\nonumber \\ \nonumber \\
v_3 \, & = & \, \frac{g_A^2 M}{16 \pi  f_\pi^2} \, \left ( \, \frac{16}{ a^2 \,
m_\pi^6}\,  - \, \frac{128}{7 a \, m_\pi^5} \, + \,\frac{16}{3 m_\pi^4} \right )
\nonumber \\ \nonumber \\
v_4 \, & = & \, \frac{g_A^2 M}{16 \pi  f_\pi^2} \, \left
( \, -\frac{256}{5 a^2 \, 
m_\pi^8}\,  + \, \frac{512}{9 a \,m_\pi^7} \, - \, \frac{16}{ m_\pi^6} \right )
 \nonumber \\  \nonumber \\
& \ldots &
\label{vi}
\end{eqnarray}
 
The expressions for the $v_i$ coefficients in eq.~(\ref{vi}) are low energy 
theorems.  As is
clear from our derivation, they are valid provided $Q/\Lambda$ is
sufficiently small.   As stated in the Introduction, one sees that 
\begin{equation}
v_j \sim \frac{1}{Q^{2 j -2}} 
\end{equation}
Moreover, it is clear that each term in the equations for all of the terms in the 
expressions for the $v_j$ 
go as $1/m_\pi^n$ with $n \ge 2$.  Thus the $v_j$ all diverge as the
chiral limit is approached. 

\section{Low Energy Theorems for ${}^3S_1$ and the ${}^3D_1$ Mixing}

We can also derive low energy theorems which test  the pionic
contributions in the
mixing between ${}^3S_1$ and the ${}^3D_1$.  Conventionally the S matrix in
this mixed channel is parameterized in the form 
\begin{equation}
S \, = \, \left( \begin{array}{l l} \cos(2 \epsilon) \, e^{ i (2 \delta_0)} \;&
\sin(2 \epsilon) \, e^{i (\delta_0 + \delta_2)}\;  \\ 
\sin(2 \epsilon) \, e^{i (\delta_0 + \delta_2)} &  \cos(2 \epsilon) \, e^{ i (
2\delta_0 )}
\end{array} \right )
\end{equation}
The leading order contribution\cite{KSW} to the $\epsilon$ parameter in $Q$
counting is 
 ${\cal O}(Q^1)$; when pions are integrated out, however, the leading
 contribution is  ${\cal O}(Q^2)$.  Thus the $\epsilon$ parameter
provides a sensitive test of the pion physics, in the sense discussed in the
Introduction.

Rather than consider the $\epsilon$ parameter globally as a function
of $k$, it is illuminating to expand the $\epsilon$ as a function of $k$:
\begin{equation}
\epsilon (p) \, = \, g_1 \, k^3 \, + g_2 \, k^5 \, + \, g_3 k^7 \, + \ldots
\label{gi}\end{equation}

In a manner analogous to the derivation of the theorems for $v_i$ we derive
theorems for the $g_i$ coefficients:
\begin{eqnarray}
g_1 \, &  = & \, \frac{\sqrt{2} g_A^2 M}{\pi f_\pi^2} \, 
\left [ \frac{-1}{8  m_\pi^2} \, + \, \frac{a}{15 m_\pi} \right ] \nonumber \\
\nonumber \\
g_2 \, & = & \, \frac{\sqrt{2} g_A^2 M}{\pi f_\pi^2} \, 
\left[ \frac{5}{12m_\pi^4} \, - \,  \frac{8 a}{35 m_\pi^3} \, + \,
 \frac{a^2}{16 m_\pi^2} \, - \, \frac{a^3}{30 m_\pi} \right ] \nonumber \\ \nonumber \\
g_3 \, & = & \, \frac{\sqrt{2} g_A^2 M}{\pi f_\pi^2} \, 
\left [ \frac{-7}{5 m_\pi^6} \, + \, \frac{16 a}{21 m_\pi^5} \, - \, 
\frac{5 a^2}{24 m_\pi^4} \, + \,  \frac{4 a^3}{35 m_\pi^3} \, - \,
 \frac{3 a^4}{64 m_\pi^2} \, + \, \frac{a^5}{40 m_\pi} \right] \nonumber\\
 \ldots
 \label{LETg}\end{eqnarray} 
The $g_j$ coefficients are also non-analytic in $Q$:
\begin{equation}
g_j \sim \frac{1}{Q^{2 j}}
\end{equation}
As with the $v_j$,  each term in the leading order expression  for $g_j$  
diverges in the chiral limit of $m_\pi \rightarrow 0$.
 
\section{Comparison With Scattering Data}

The $v_j$ and $g_j$ coefficients are observables which can be
extracted from the nucleon-nucleon scattering data.  We have extracted
these coefficients from the scattering data as parameterized in the
Nijmegen\cite{Nij} partial wave analysis.  The extracted coefficients
for the $v_j$ in the triplet channel are taken from
ref. \cite{Stoksa}.  For the $v_j$ in the singlet channel we have done
a least squares fit directly from the Nijmegen phase shifts at very
low energies.  Our fits in the singlet channel agree quite well with
fits of the coefficients calculated from potential models fitted to
the phase shift data \cite{Stoksb}.  We have also extracted the $g_i$
coefficients using a least squares fit to the low energy data.  In
Table~(\ref{comp}) we compare the extracted coefficients with those
predicted by the low energy theorems.

It is apparent from Table~(\ref{comp}) that the low energy theorems for the
$v_i$ coefficients fail quite badly in both the singlet and triplet channels.
In all cases they predict coefficients  which are several times the extracted
ones, typically by a factor of $\sim 5$.  The $v_2$ in the triplet channel has a more
spectacular failure---overstating the extracted value by more than a factor of
20. The low energy theorems for the $g_i$ coefficients also 
do rather poorly.   The predicted
$g_i$ coefficients are all  more than 100\%
 greater than those extracted from the
data, essentially meaning no prediction.  The coefficient $g_2$ is $\sim 4$
times that extracted from the data while predicted $g_3$  is $\sim 8$ the 
extracted one.  

\section{Discussion}

The clear failure of the chiral expansion to describe the $v_i$
coefficients for the s-wave may seem at first glance quite surprising. 
After all, 
Kaplan, Savage and  Wise\cite{KSW} described s-wave scattering using the PDS 
approach and, using a global fit for the parameters, appear to successfully 
describe the phase shifts up to $k \sim 300$ MeV. 
 How, then, can the $v_i$ coefficients in the 
effective range expansion describing the same scattering data
all be badly wrong?  The answer is, in fact, quite simple. 
 The principal point is simply that the 
general shape of the phase shift curve---a very rapid rise at low $k$ followed
by a slow decrease is implied directly from the effective range expansion with 
large $a$; {\it any} theory with free parameters consistent with the effective range
expansion will be able to reproduce the crude shape. Thus the
ability to crudely reproduce the shape of the data is not a stringent test.
 Moreover, 
the {\it global} fit  used  in ref. \cite{KSW}  masks possible subtle problems with
the detailed shape of the fit to the phase shifts.  
However, there is an important issue,  related to the quality of the fit of the s-wave
scattering in ref. \cite{KSW} which goes to the heart of the issue relating to
the role of pions in
the effective field theory program.

As we have argued in the Introduction, the easiest way to 
test the role of pions
in the  EFT is to compare a calculation in an EFT with explicit pions to
a calculation at the same order in an EFT with the pions integrated out. 
If the chiral expansion is working well we should find that the inclusion of
explicit pions both substantially improves the quality of agreement of
 the low $k$ observables with the data and increase  the range in $k$ for which
 the EFT is useful.
Here, since we are working at order $Q^2$ in $k \cot (\delta)$, the theory with
pions integrated out is simply the effective range expansion up to 
the effective range term.  

\begin{figure}[p]
\epsfig{figure=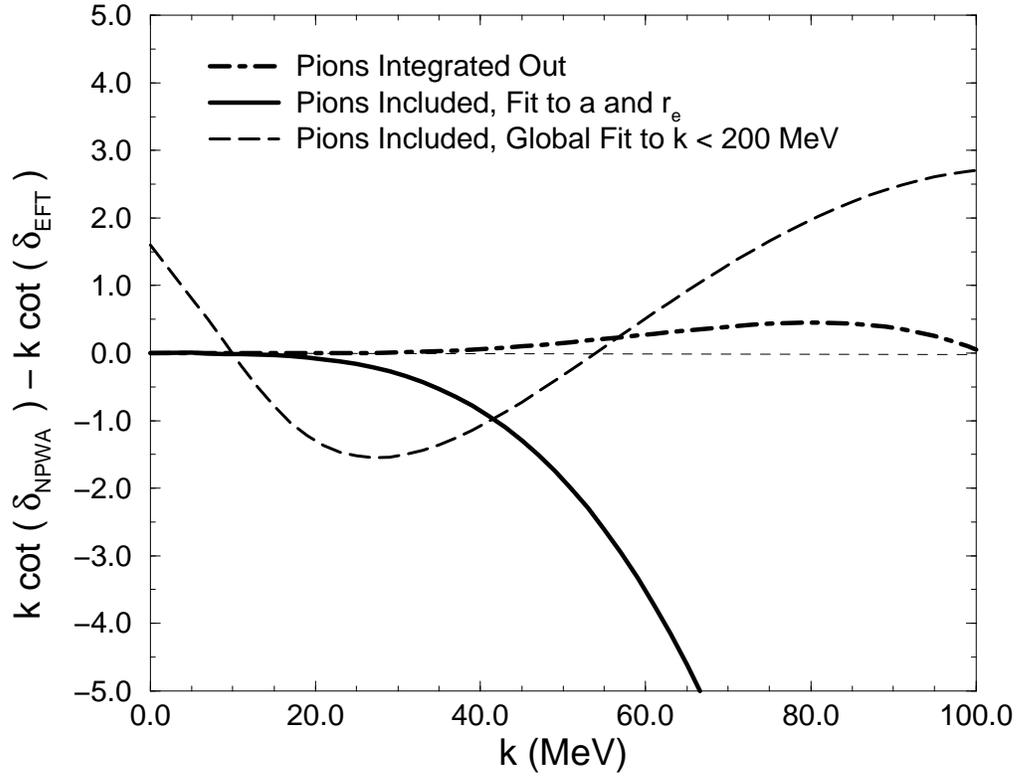, height=12cm}
\caption{
\label{fig1}
$k \cot(\delta_{\rm EFT}) - k \cot
 (\delta_{\rm NPWA})$ versus $k$ where the subscript NPWA indicates  the Nijmegen partial wave analysis and the subscript EFT indicates the effective field theory. }

\end{figure}

In fig.~\ref{fig1} we compare the difference of
$k \cot (\delta)$ from the Nijmegen\cite{Nij} partial wave analysis in the ${}^1S_0$  channel
from three order $Q^2$ EFT predictions: a) a pions-integrated-out theory 
 with parameters
fit to the observed $a$ and $r_e$ (this is precisely the first two terms in
the effective range
expansion);  b) a pions-explicitly-included theory with
the parameters in eq.~(\ref{kcotd2}) fit to the observed $a$ and $r_e$ using
eq.~(\ref{kcotd1}); c) a
pions-explicitly-included theory
from  the global fit  of Kaplan, Savage and Wise\cite{KSW}.

We have plotted the difference from the data to focus attention on the accuracy
of the predictions (in the spirit of refs. \cite{steelea,steeleb}).  Plotting
the data in this manner also removes the visual effect of reproducing the rise and
fall of the phase shifts.
 The striking thing about fig.~\ref{fig1} is that the 
inclusion of pions 
does not improve the quality of the fit at low $k$ compared to the pion-less 
theory.  Indeed, at moderately low energies, the inclusion of pions markedly worsens the agreement
of theory with experiment compared to the pion-integrated-out theory. This 
strongly suggests that the chiral expansion is not under control. 
Having seen this, it is not
so surprising that the global fit is so poor at low energies; it was designed
to compromise the low energy behavior in order to do better at larger $Q$.  In
testing  low energy observables, however, it is more appropriate to 
use coefficients fit to the low energy behavior.  Indeed,  there is an
ambiguity in using the global fit parameters. 
In the plot  we have evaluated $k$ times the cotangent
of the phase shifts from the
expression in ref.\cite{KSW}.  Alternatively,  
we could have used  the global fit parameters
directly in the expression for $k \cot{\delta}$ in eqs.~(\ref{kcotd1}) and
(\ref{equiv}).
While these two forms agree to order $Q^2$, they differ at higher order.
As the fit was done using $\delta$ rather than $k \cot (\delta)$
we felt it more reasonable to plot it this way.  Had we used the same
parameters in the $k \cot (\delta)$ expression the fit   would have markedly 
worsened.

The $v_i$  coefficients directly  measure the improvement  of the theory with 
explicit pions  over the theory with pions integrated out. 
The pion-less theory at
this order is simply the effective range expansion up to the $r_e$ 
term; the  $v_i$ coefficients parameterize the way in which the phase shifts
deviate from this expression.   Thus the failure of the effective field theory 
treatment to get the $v_i$ coefficients is also evident in fig.~\ref{fig1}
and suggests a failure of the ``including
pions'' part  of the $Q$ counting
program---at least so far as the  s-wave scattering is concerned.

It should also be remarked that the failure in the prediction of the
$\epsilon$ parameter evident in these $g_i$ coefficients is much worse than
suggested by the plot of $\epsilon$ against $k$ in ref. \cite{KSW}.  
In  ref. \cite{KSW}, at low $k$, the predicted $\epsilon$ appears to be 
$\sim$ 40\%  above the data at low energy 
(although the predicted shape differs greatly from the observed one at higher
energies).  How can this be reconciled with a $g_1$ parameter predicted in our
low energy theorem which is 2.3 times the observed one?  The origin of this
discrepancy is related to the fit used.  Using the global fit parameters of
ref. \cite{KSW} and directly calculating  $g_1$ yields a coefficient which is
only $\sim 40$\% greater than that extracted from the data. 
In contrast, using the physical  scattering length, as we do in 
eq.~(\ref{LETg}), we obtain one which is $\sim$ 130 \% larger.
As both the calculation using the physical $a$ and the global fit are at the
same order in $Q$, one must view the better agreement of the calculation using 
the global fit parameters as being fortuitous.  Moreover, the large disagreement 
between calculation based on the global fit and the one based on the $a$ is
itself an indication that  the expansion has broken down.

It should be obvious from the pattern of failure for both of these 
low energy theorems
that the  $\Lambda$, the scale of the
short distance physics, is {\it not} much larger than $m_\pi$.  
 The derivation in ref.~\cite{CH} provides some
insight as to why this might be expected
 which may not be as immediately apparent in
the PDS formalism. 
In the derivation using the cutoff formalism, the expansion in $m_\pi/\Lambda$
came in two distinct ways.  The first was in justifying the perturbative
inclusion of the pion exchange interaction in a manner similar to the Born series but with boundary
conditions at some  matching radius $R$ to reflect the short distance physics.
The second  was in 
the expansion of $m_\pi R$ in the various integrals which arise.
For this expansion to be valid we must take $m_\pi R \ll 1$. 
A central issue is whether
this is true in practice.  Of course, the value of R is somewhat arbitrary 
having  been introduced
by us.  We are not free, however, to make $R$ arbitrarily small; by
construction, $R$ must be large enough so that the effects
of the short distance  potential on the wave function are contained within $R$
with an accuracy comparable to the order at which we are working.  

In fact,
comparisons with potential models suggest that $ m_\pi R\sim 1$, {\it i.e.} the 
range of the short distance potential is comparable to $1/m_\pi$.  
One rather compelling way to see this is found in sect. IV of Scaldeferri
{\it et al} \cite{Scal}.  There, using a generalization of conventional
 effective range
theory,   it is shown that for {\it any}
energy independent  potential model 
consisting of a one pion exchange plus a  short distance potential
(including a nonlocal potential), there
is a rigorous lower bound on $R$.  In particular, it is shown that in order 
to fit the scattering length and effective range, the short range potential must make 
a non-negligible contribution to the wave function out to at least 1.1 fm---
{\it i.e.} $ R >1.1$ fm.  This corresponds to $m_\pi R >  .78$.  Moreover, it
is clear from the derivation that this bound on $R$ cannot be saturated, since the
bound is saturated only if the wave function is strictly zero for $r<R$; the actual
value of $R$ beyond which the short distance potential ceases to make
significant contributions is accordingly expected to be significantly larger
suggesting a value of $m_\pi R \sim 1$.  
For example, if the short distance potential for the ${}^1S_0$ channel is taken to be a
square well whose depth and range is fit to the scattering length and effective range,
then  $R=2.3$ fm \cite{Scal}, which corresponds to  $m_\pi R \approx 1.6$.
  An obvious conclusion is that
the scale of the short distance contribution to the nucleon-nucleon interaction
is not well separated from $m_\pi$ making the chiral expansion highly
problematic. 

Of course, it might be argued that 
the derivation in ref. \cite{Scal} includes only one pion exchange
and short distance physics.  It does not  include higher order 
long distance effects such
as two pion exchange which could alter the result.  However, if one is  in 
regime where $Q$ counting is valid, then two pion exchange and similar mechanisms
only contribute to $k \cot (\delta)$ at order $Q^3$ or higher and  may be
neglected, implying that the bound in ref. \cite{Scal} is valid up to corrections of
order $Q/\Lambda$.  
  
The justification of the perturbative, Born-like expansion for the inclusion
of the pion exchange interaction  also may be problematic.
In ref. \cite{KSW} the 
quantity,  $16 \pi f_\pi^2/(g_A^2 M)$, was taken to play  the role of $\Lambda$
and the dimensionless parameter, $\eta = ( m_\pi g_A^2 M) /(16 \pi f_\pi^2)$, was
identified as playing the role of the expansion parameter.  In terms of the
derivation of ref. \cite{CH}, it is clear that $\eta = f_1 (k^2=0)$.
The potential difficulty for $\chi$ PT was identified in ref. \cite{KSW}---
$\eta$ is rather large; numerically $\eta \approx .47$.  If the
relevant parameter is in fact $\eta$, one might expect a slowly converging theory.  On
the other hand, if the relevant parameter is more like 2 $\eta$, things are
quite out of control. Moreover, this expansion parameter arises in the context
of the central force.  In the triplet channel there is also a tensor force. 
While formally in $Q$ counting the effect of the tensor force is higher order,
experience in the nuclear physics problem is that it is quite important owing
to large numerical factors.  Given both the problems with the $m_\pi R$
expansion and the justification of a perturbative treatment for the pion,
the failures of the low energy theorems are not so surprising.

 The possibility of problems with the perturbative treatment of the pion 
 exchange interaction implicit in the PDS scheme has been raised previously by
Gegelia\cite{Gegelia2} and Steele and Furnstahl \cite{steeleb} 
using rather different arguments.  Here
 we have shown explicitly in the 
case of the $v_i$ coefficients in s-wave scattering and the $g_i$ coefficients
in the ${}^3 S_1$--${}^3 D_1$ mixing that the``including pions''
aspect of the systematic $Q$ counting  approach in nuclear physics fails
badly.  We have also given  general arguments that there exist ``short
distance'' scales in the nucleon-nucleon problem which are $\sim m_\pi$
suggesting that the ``including pion'' parts of $Q$ counting may be expected
to have problems.  It remains to be seen just how widespread 
these problems will turn out to be.   Clearly the perturbative pions work well
in high partial waves, but the question remains whether observables sensitive 
to both 
the pion physics and to s-wave nucleon-nucleon interactions are generally well
 described in $Q$ counting.
 In order to determine this,
it is necessary   to study a number of  
observables which  are  particularly sensitive to  pion range physics.

The authors thank Silas Beane and Daniel Phillips for interesting
discussions.  We are indebted to Vincent Stoks for providing us with numerical
values of the effective range expansion coefficients for potential model
calculations of the ${}^1S_0$ channel.  We also thank Mike Birse for 
pointing out to us the issue discussed in the appendix.  
TDC gratefully acknowledges the support
of the U.S. D.O.E.

\appendix
\section{The Chiral Limit and the Equivalence of the Cutoff and PDS Approaches}

In ref. \cite{CH} it was shown that the PDS and cutoff
 expressions for $k \cot {\delta}$ 
are equivalent provided the coefficients in the two approaches are
matched as given by eq.~(\ref{equiv}). 
 However, as pointed out by Birse\cite{Birse} there is an apparent difficulty with
this equivalence as the chiral limit is approached.  Note that $Q$ counting formally holds
if both $1/a$ and $m_\pi$ are small, however the combination $m_\pi a$ is
unconstrained and can take any value.  Thus, a scheme which
implements $Q$ consistently   should be able to correctly describe the formal 
limit $m_\pi \rightarrow 0$, $1/a \rightarrow 0$, $m_\pi a \rightarrow 0$.
This formal limit, is of course nothing but the chiral limit in a regime where
$Q$ counting holds.
We note, at the outset, that this limit is clearly not of immediate practical
concern since in nature $m_\pi a \gg 1$.  Never-the-less,  if the $Q$ counting scheme is
viable, nothing in the formalism should prevent an  approach to the 
chiral limit.  The potential problem is quite clear in the last equation of 
eqs.~(\ref{equiv}): the left-hand side goes to zero in the chiral limit
while the right-hand side does not, suggesting that the equivalence cannot
hold for all values of $m_\pi$.

The difficulty stems from a minor inconsistency  with the PDS 
formulation which can  be cured easily and which has no observable
consequences for NN scattering.   Note that as the chiral limit of the
last equation of eqs.~(\ref{equiv}) is taken, the right-hand side is
not merely nonzero, it is $\mu$ dependent.  This is a direct consequence of
the fact that in the formulation of PDS in ref.\cite{KSW}, the scattering
length in the chiral limit is not renormalization group invariant.  The PDS 
expression for  $k \cot (\delta)$ at order $Q^2$ is \cite{CH}
\begin{eqnarray}
k \cot (\delta) \, & = &\, -\frac{1}{a_0} \, + \, 
\left [ \frac{g_A^2  M}{16 \pi f_\pi^2} \, \left ( m_\pi^2 \,   \ln
\left (\frac{m_\pi}{\mu} \right ) \, - \,
 m_\pi^2 \, + \, \frac{1}{a_0^2} \, - \, 2 \frac{\mu}{a_0}  \, + \mu^2 \right )
\, + \, \frac{D_2 M}{4 \pi} \left ( m_\pi^2 \mu^2 \, -  \, 
\frac{2 m_\pi^2 \mu}{a_0} \, + \, \frac{m_\pi^2}{a_0^2} \right)  \right ] 
\nonumber \\ \nonumber\\
 \, & + & \, k^2  \, \left \{\frac{C_2 M}{4 \pi} \, \left ( \mu^2 \, - \, 
 \frac{2 \mu}{a_0} \, + \, \frac{1}{a_0^2} \right ) \right \} 
\, -  \,  \frac{1}{a_0^2} \, \frac{g_A^2  M}{64 \pi f_\pi^2} \,
\left( \frac{m_\pi^2}{k^2} \right ) 
\ln \left
(1 + \frac{4 k^2}{m_\pi^2} \right ) \nonumber \\ \nonumber\\
 \, & +  & \, \frac{m_\pi}{a_0} \, \frac{g_A^2  M}{16 \pi f_\pi^2} \, 
 \left( \frac{m_\pi}{k} \right )\,
 \tan^{-1} \left ( \frac{2 k}{m_\pi} \right ) \,
 + \, m_\pi^2 \, \frac{g_A^2  M}{64 \pi f_\pi^2} \, 
 \ln \left (1 + \frac{4 k^2}{m_\pi^2} \right ) 
 \label {kcotdpds}\end{eqnarray}
Taking the limit $k \rightarrow 0$ picks up $-1/a$ and subsequently taking the chiral limit
yields:
\begin{equation}
-\left . \frac{1}{a} \right |_{m_\pi = 0} \, = \, -\frac{1}{a_0} \, + \, 
\frac{g_A^2  M}{16 \pi f_\pi^2} \, \left (  - \, 2 \frac{\mu}{a_0} 
 \, + \mu^2 \right )
\label{chilim}\end{equation}
which depends on $\mu$ and thus is manifestly not renormalization group invariant.
Note,  this problem is not immediately apparent in ref. \cite{KSW} since
the expression for the scattering length is only given after $\mu$ is set
equal to $m_\pi$.

The reason for the $\mu $ dependence of eq.~(\ref{chilim}) is clear.  No counter
term has been given in ref. \cite{KSW} which can absorb it.  On the other hand,
this can be  dealt with easily.  The simplest way to rewrite the coefficient $C_0$
as $C_0^{\rm nonpertub} + C_0^{\rm perturb}$ where $C_0^{\rm nonpertub}$
is iterated to all orders and contributes to $k \cot (\delta )$ at order $Q$
and reproduces the physics $1/a_0$.  In contrast $C_0^{\rm perturb}$ serves
as a counter term absorbing the $\mu$ dependence, is not iterated to all orders
and only contributes at order $Q^2$.  Note, however that in calculating cross-sections
$C_0^{\rm perturb}$ and $D m_\pi^2$ enter in exactly the same way.  Thus, only
the combinaton $C_0^{\rm perturb} + D m_\pi^2$ affects the scattering amplitude.
To the extent that $D$ was fit directly from scattering data one can include
$C_0^{\rm perturb}$  by making the subsititution
$D^{\rm fit} m_\pi^2 \rightarrow D m_\pi^2 + C_0^{\rm perturb}$ with {\it no 
change} to the predicted scattering.  Thus, the only effect of
including $C_0^{\rm perturb}$ as a separate counter term is on the behavior as
the chiral limit is approached.  

After the work in this appendix was finished, we became aware of the work
of Mehen and Stewart\cite{MS} in which substantially the same conclusions
are reached, namely that $\mu$ independence requires splitting a perturbative
contribtuion from $C_0$ and that only the contribution from the combination of the $D$ plus the perturbative part of $C_0$ contributes in nucleon-nucleon scattering.  We note that the work
of Mehen and Stewart is far more general than this and proposes a more transparent way to implement dimensional regularization than PDS.

\newpage

\begin{table}[p]
\begin{tabular}{||c| c | c | c ||}
$\delta$ (${}^1 S_0$ channel) &  $v_2$ (${\rm fm}^{3})$ & $v_3$ (${\rm fm}^{5})$ & $v_4$ (${\rm
fm}^{7})$\\ \hline \hline & & &\\
low energy theorem &  -3.3 & 17.8 & -108. \\ \hline & & & \\
partial wave analysis& -.48 & 3.8 & -17. \\ 
\end{tabular}

\begin{tabular}{||c| c | c | c ||}
$\delta$ (${}^3S_1$  channel)&$v_2$ (${\rm fm}^{3})$ & $v_3$ (${\rm fm}^{5})$ & $v_4$ (${\rm
fm}^{7})$\\ \hline \hline & & & \\
low energy theorem &  -.95 & 4.6  & -25.  \\  \hline & & & \\
partial wave analysis & .04 & .67 &  -4.0\\ 
 \end{tabular} 
 
 \begin{tabular}{||c| c | c | c ||}
$\epsilon$ (${}^3S_1$--${}^3D_1$  mixing) &$g_1$ (${\rm fm}^{3})$ & $g_2$ (${\rm fm}^{5})$ &
$g_3$ (${\rm
fm}^{7})$\\ \hline \hline
& & &\\
low energy theorem &  3.9 & -86.  & 1.8 $10^3$  \\  \hline
& & & \\
partial wave analysis & 1.7 & -26.  &  2.2 $10^2$\\ 
 \end{tabular}

\caption{A comparison of the predicted effective range expansion 
coefficients, $v_i$,   for the ${}^1S_0$ 
and ${}^3S_1$ 
channels and the predicted  $g_i$ coefficients 
in the expansion of $\epsilon$ with coefficients extracted
from the Nijmegen partial wave analysis.}
\label{comp}
\end{table}

\end{document}